\PassOptionsToPackage{table}{xcolor} 

\newif\ifarxivmode
\arxivmodetrue 

\ifarxivmode
	\documentclass[conference,year=24]{IEEEtran}
\else
	\documentclass[conference,year=24]{fmcad}
\fi

\newif\iffinal
\finalfalse

\usepackage{calc}
\usepackage[framemethod=TikZ]{mdframed}
\usepackage{forest}
\usepackage{subcaption}
\usepackage{xspace}
\usepackage{courier}
\usepackage{makecell}
\usepackage{svg}
\usepackage{booktabs,makecell}
\ifarxivmode
	\usepackage[hidelinks]{hyperref}
\fi
\usepackage[altpo,epsilon]{backnaur}
\usepackage{amsmath}
\usepackage[capitalise]{cleveref}
\usepackage[amsthm,thmmarks,amsmath]{ntheorem}
\usepackage{mathtools}
\usepackage{caption}
\usepackage{textcomp}  
\usepackage{multirow}
\usepackage{paralist}
\usepackage{balance}
\usepackage{dirtytalk}
\usepackage{afterpage}
\usepackage{adjustbox} 
\usepackage{backnaur}
\usepackage{tikz}
\usetikzlibrary{tikzmark,calc,shapes, positioning,arrows.meta}
\newtheorem{theorem}{Theorem}
\crefname{theorem}{Thm.}{Thm.}

\usepackage{enumitem,amssymb} 
\newlist{todolist}{itemize}{2}
\setlist[todolist]{label=$\square$}

\usepackage{listings,array,varwidth,listings-rust}
\usepackage{lstautogobble}  
\usepackage{color}          
\usepackage{zi4}            
\definecolor{darkgreen}{RGB}{0, 150, 0}
\definecolor{bluekeywords}{rgb}{0.13, 0.13, 1}
\definecolor{greencomments}{rgb}{0, 0.5, 0}
\definecolor{redstrings}{rgb}{0.9, 0, 0}
\definecolor{graynumbers}{rgb}{0.5, 0.5, 0.5}
\definecolor{pastelpink}{RGB}{255, 182, 193}
\definecolor{pastelblue}{RGB}{173, 216, 230}
\definecolor{pastelgreen}{RGB}{182, 255, 182}
\definecolor{pastelyellow}{RGB}{255, 255, 204}
\definecolor{pastelpurple}{RGB}{216, 191, 216}
\definecolor{pastelmustard}{RGB}{255, 219, 88}
\usepackage{listings}
\newcommand{\bmc}{\text{BMC}\xspace}

\newcommand{\seahorn}{\textsc{SeaHorn}\xspace}
\newcommand{\seabmc}{\textsc{SeaBMC}\xspace}

\newcommand{\llvm}{\textsc{LLVM}\xspace}

\newcommand{\smt}{\textsc{SMT}\xspace}
\newcommand{\smtlib}{\textsc{SMT-LIB}\xspace}

\newcommand{\seair}{\textsc{SEA-IR}\xspace}
\newcommand{\oseair}{\textsc{OSEA-IR}\xspace}

\newcommand{\mbedtls}{\texttt{mbedTLS}\xspace}
\newcommand{\rust}{\text{Rust}\xspace}

\newlength{\markoffset}
\iffinal
	\newcommand{\ag}[1]{\textcolor{purple}{}\xspace}
	\newcommand{\sid}[1]{\textcolor{darkgreen}{}\xspace}

\else
	\newcommand{\ag}[1]{\textcolor{purple}{[AG: #1]}\xspace}
	\newcommand{\sid}[1]{\textcolor{darkgreen}{[SP: #1]}\xspace}
\fi

\newcommand{\symfun}{\ensuremath{sym}\xspace}
\newcommand{\sym}[1]{\ensuremath{sym( #1 )}}

\newcommand{\bv}[1]{\ensuremath{\mathit{bv}( #1 )}}

\newcommand{\code}[1]{\lstinline[basicstyle=\footnotesize\ttfamily,style=SeaC++] {#1}\xspace}
\newcommand{\tcode}[1]{\text{\code{#1}}}

\newcommand{\bparagraph}[1]{\noindent\textbf{#1}}

\newenvironment{bnfsplit}[1][1.2\textwidth]
{\minipage[t]{#1}$}
{$\endminipage}

\newcommand\sdef{\ensuremath{\triangleq}}

\newcommand{\entangle}{\ensuremath{tngle}\xspace}

\newcommand{\val}{\ensuremath{val}\xspace}
\newcommand{\ret}[1]{\ensuremath{ret#1}\xspace}
\newcommand{\loc}{\ensuremath{addr}\xspace}

\newcommand{\marktext}[2][pastelmustard]{
	\hspace{\markoffset}\colorbox{#1}{\makebox[\widthof{#2}][l]{\color{black}\vphantom{#2}#2}\hspace{-\fboxsep}}%
}

\newcommand{\Zrepeat}[2]{\foreach \n in {1, ..., #1}{#2}}
\newcommand{\sind}[1][1]{\;\Zrepeat{#1}{\;}}

\lstdefinelanguage{seair}%
{morekeywords={abstract,%
			assume,sassert,assert,phi,br,%
			alloca,malloc,load,store,%
			gep,halt,select,%
			isderef,isalloc,ismod,resetmod,%
			fun,main,add,ptoi,itop,free,memmove,memcpy,%
			mk_own,mut_mkbor,mut_mksuc,die,%
		},%
	mathescape=true,%
	sensitive,%
	commentstyle=\color{blue},
	morecomment=[l]//,%
	morecomment=[s]{/*}{*/},%
	morestring=[b]",%
	morestring=[b]',%
	tabsize=2,
	captionpos=b,
	showstringspaces=false
}[keywords,comments,strings]%

\lstdefinestyle{SeaC++}{
	language=C++,
	morekeywords={size_t,assume, sassert},
}
\lstdefinestyle{SeaC}{
	language=C,
	morekeywords={size_t,assume, sassert},
}

\newcommand{\drawfatptr}[5] {
	\begin{tabular}{|c|c|c}
		\cellcolor{purple!30}$\mathit{ptr}$ & \cellcolor{purple!30} $\mathit{#2}$ & \cellcolor{purple!30} #4          \\
		\cellcolor{pink!60} $\mathit{#1}$   & \cellcolor{pink!60} $\mathit{#3}$   & \cellcolor{pink!60} $\mathit{#5}$
	\end{tabular}
}

\newcommand{\mutborbor}{\code{mut_mkbor}}
\newcommand{\mutborsuc}{\code{mut_mksuc}}

\newcommand{\mkown}{\code{mk_own}}

\newcommand{\setcache}{\code{set_cache}}
\newcommand{\getcache}{\code{get_cache}}

\newcommand{\die}{\code{die}}
\newcommand{\cpyone}{\code{cpy_mkcpy1}}
\newcommand{\cpytwo}{\code{cpy_mkcpy2}}
\newcommand{\beginunique}{\code{begin_unique}}
\newcommand{\uniqueend}{\code{end_unique}}

\newcommand{\machzero}{\ensuremath{\mathcal{M}_{0}}\xspace}
\newcommand{\machone}{\ensuremath{\mathcal{M}_{1}}\xspace}

\newcommand{\machtwo}{\ensuremath{\mathcal{M}_{2}}\xspace}

\newcommand{\machthree}{\ensuremath{\mathcal{M}_{3}}\xspace}

\newlist{customitemize}{itemize}{1}
\setlist[customitemize]{label={\textbf{Custom Label}}}

\usepackage{graphicx}

\begin{document}
\title{Ownership in low-level intermediate representation}
\ifarxivmode
	\author{\IEEEauthorblockN{Siddharth Priya\IEEEauthorrefmark{1}}
		\IEEEauthorblockA{\textit{siddharth.priya@uwaterloo.ca}} \and
		\IEEEauthorblockN{Arie Gurfinkel\IEEEauthorrefmark{1}}
		\IEEEauthorblockA{\textit{arie.gurfinkel@uwaterloo.ca}}
	}
\else
	\author{\IEEEauthorblockN{Siddharth Priya\orcid{0000-0002-2172-9525}\IEEEauthorrefmark{1}}
		\IEEEauthorblockA{\textit{siddharth.priya@uwaterloo.ca}} \and
		\IEEEauthorblockN{Arie Gurfinkel\orcid{0000-0002-5964-6792}\IEEEauthorrefmark{1}}
		\IEEEauthorblockA{\textit{arie.gurfinkel@uwaterloo.ca}}
	}
\fi

\pagestyle{plain}
\maketitle              
\begin{abstract}
	The concept of ownership in high level languages can aid both the programmer and the compiler to reason about the validity of memory operations.
	Previously, ownership semantics has been used successfully in high level automatic program verification to model a reference to data by a first order logic (FOL) representation of data instead of maintaining an address map.
	However, ownership semantics is not used in low-level program verification.
	We have identified two challenges.
	First, ownership information is lost when a program is compiled to a low-level intermediate representation (e.g., in \llvm IR).
	Second, pointers in low-level programs point to bytes using an address map (e.g., in unsafe \rust) and thus the verification condition (VC) cannot always replace a pointer by its FOL abstraction.
	To remedy the situation, we develop ownership semantics for an \llvm-like low-level intermediate representation.
	Using these semantics, the VC can opportunistically model some memory accesses by a direct access of a pointer \emph{cache} that stores byte representation of data.
	This scheme reduces instances where an address map must be maintained, especially for mostly safe programs that follow ownership semantics.
	For unsafe functionality, memory accesses are modelled by operations on an address map and we provide mechanisms to keep the address map and pointer cache in-sync.
	We implement these semantics in \seabmc, a bit-precise bounded model checker for \llvm.
	For evaluation, the source programs are assumed to be written in C.
	Since C does not have ownership built-in, suitable macros are added that introduce and preserve ownership during translation to \llvm-like IR for verification.
	This approach is evaluated on mature open source C code.
	For both handcrafted benchmarks and practical programs, we observe a speedup of 1.3x--5x during \smt solving.
\end{abstract}
\section{Introduction}
\label{sec:intro}
Ownership is a scheme to control aliasing of references in high level languages.
It has been studied in a long line of academic research~\cite{DBLP:conf/pldi/FahndrichD02,DBLP:conf/pldi/GrossmanMJHWC02,DBLP:journals/toplas/MatsushitaTK21,DBLP:conf/ecoop/NobleVP98}.
More recently, the concept has gained attention due to \rust, a popular systems language that offers low level control like C/C++ and uses ownership semantics to record aliases and mutation of data.
In \rust, 
\begin{inparaenum}[(1)]
  \item a value has exactly one owner,
  \item a reference to a value (called a \emph{borrow}) cannot outlive the owner, and
  \item a value can have one mutable reference \emph{xor} many immutable references.
\end{inparaenum}
A program that follows this programming discipline allows the \rust compiler to reason about memory safety statically.
However, for reasons of expressivity and performance, programs may need to break this discipline for certain operations.
For this, \rust provides \emph{unsafe} code blocks where the static checks are temporarily turned off.

While, ownership can aid in generating correct and efficient code,
it is also useful in program verification.
Usually, the presence of aliasing necessitates an \emph{address map} to soundly model object accesses through different aliases.
With ownership, this map can be eliminated when it is known that only a single reference exists.
This has been useful for program verification.
For example, the Move Prover~\cite{DBLP:conf/tacas/DillGPQXZ22} replaces references by objects in the generated verification conditions (VC). Similarly, RustHorn~\cite{DBLP:journals/toplas/MatsushitaTK21,DBLP:conf/pldi/MatsushitaDJD22} is able to generate pure First Order Logic VC for safe \rust programs without introducing a memory model.

The advances in verification using ownership semantics have not made their way to verification of low-level programs.
One of the problems is that low-level languages do not support ownership out of the box.
As an example, \llvm bitcode is a register based intermediate representation (IR) used by C, C++, and, \rust compilers.
It only has an attribute for marking pointers as \code{noalias}~\cite{NoAlias-LLVM} and no ownership operations.
The \code{noalias} attribute is useful for optimization.
However, the semantics of \code{noalias} in unclear and has caused confusion~\cite{NoAlias-Rust}.
Another challenge is that ownership in high-level language does not translate directly to low-level settings.
For example, in verification of safe \rust programs, it is correct to model a reference by the FOL representation of the value it refers to.
However, this model does not work for \llvm-like IR (and unsafe \rust) because such languages (dialect) have pointers that treat values as a collection of bytes and rely on pointer arithmetic to access individual bytes.
In verification, the standard solution  models memory using an address map from addresses to byte or word values.
However, such address maps are expensive to execute symbolically.

This work improves the state-of-the-art by the following contributions.
First, we develop an ownership semantics for an \llvm-like low level language that operates on single words in memory.
This language replaces unrestricted aliasing with mutable borrow, read-only borrow, and copy operations that track outstanding aliases for a memory allocation.
Second, we define a caching mechanism for capturing data at a pointer itself. This cache can by written and read by operations on the pointer.
A pointer cache allows us to replace memory accesses in the generated VC by the cache whenever correct to do so.
This can simplify the VC and improve the solving time.
For mostly safe programs, many memory accesses may be replaced by pointer cache accesses.
These semantics are discussed in~\cref{sec:opsem}.

Third, we discuss our design for VCGen and especially modelling the borrow operation in~\cref{sec:vcgen}.
Borrowing temporarily transfers memory access rights from the lender pointer (a.k.a. \emph{lender}) to the borrowing pointer (a.k.a. \emph{borrower}).
In the given semantics, this means that the pointer cache is also copied from lender to borrower.
However, the borrower is assumed to return the borrow by the first instance of a memory access by a lender.
This means that the updated cache at the borrower \emph{must} be copied back to the lender before this.
This transfer ordinarily requires a memory model that allows shared accesses between the borrower and lender aliases.
A more efficient way uses prophecy variables~\cite{DBLP:journals/tcs/AbadiL91} and was first proposed in~\cite{DBLP:journals/toplas/MatsushitaTK21}.
We adapt the prophecy solution to our setting.

Fourth, to make our ownership semantics practical, we add support for multi-word memory operations and evaluate the semantics by incorporating it in \seabmc~\cite{DBLP:conf/fmcad/PriyaSBZVG22}.
\seabmc is a bit-precise bounded model checking engine for \seahorn that uses an SMT solver as its backend.
To ease writing programs using these ownership semantics, the user writes C programs laced with calls to ownership macros.
During compilation, the macros expand to \llvm intrinsics that are then interpreted using the given semantics to generate VC.
The benchmark programs are a mix of handcrafted examples and practical programs.
The handcrafted examples are used to fine tune performance and show what is possible.
The practical programs are from the \mbedtls project - an open-source SSL/TLS library and has routines for encryption and secure communications.
We get a speedup of 1.3x--5x during SMT solving.
The evidence shows that the verification simplicity (speedup) correlates positively with the number of memory accesses that can be replaced by pointer cache accesses in a program.

\section{\oseair language}
\label{sec:opsem}
In this section, we present the syntax and semantics of the \oseair language.
To simplify the presentation, we propose machines that work with a single datatype $bv(N)$, a bit-vector of $N$ bits, as the \textit{word} size.
We impose two restrictions.
First, all memory operations - \textit{allocation}, \textit{load}, and, \textit{store} work on a single word.
Second, the machine can only store integers in memory and does not support \textit{store} and \textit{load} of pointers.
We lift the first restriction in~\cref{sec:multiword},
and the second in
\ifarxivmode
	~\cref{app:appendixa}.
\else
	an extended version~\cite{priya2024ownershiplowlevelintermediaterepresentation}.
\fi

\begin{figure*}[t]
	\centering
	\begin{minipage}{0.9\linewidth}
		\scriptsize
		\begin{bnf*}
			\bnfprod{S}
			{\dots \bnfor \bnfpn{OS}} \\
			\bnfprod{RDEF}{
				\begin{bnfsplit}
					\dots
					\bnfor \\
					\bnfpn{P}, \bnfpn{M} = \texttt{mk\_own} \bnfsp \bnfpn{R}, \bnfpn{M}
					\\
					\bnfpn{P} = \texttt{mut\_mkbor} \bnfsp \bnfpn{P}
					\bnfor
					\bnfpn{P} = \texttt{mut\_mkbor\_off} \bnfsp \bnfpn{P},\ \bnfsp \bnfpn{R}
					\bnfor
					\bnfpn{P} = \texttt{mut\_mksuc} \bnfsp \bnfpn{P}
					\bnfor \\
					\bnfpn{P} = \texttt{ro\_mkbor} \bnfsp \bnfpn{P}
					\bnfor
					\bnfpn{P} = \texttt{ro\_mkbor\_off} \bnfsp \bnfpn{P},\ \bnfsp \bnfpn{R}
					\bnfor
					\bnfpn{P} = \texttt{ro\_mksuc} \bnfsp \bnfpn{P}
					\bnfor \\
					\bnfpn{P} = \texttt{cpy\_mkcpy1} \bnfsp \bnfpn{P}
					\bnfor
					\bnfpn{P} = \texttt{cpy\_mkcpy1\_off} \bnfsp \bnfpn{P},\ \bnfsp \bnfpn{R}
					\bnfor
					\bnfpn{P} = \texttt{cpy\_mkcpy2} \bnfsp \bnfpn{P}
					\bnfor 
					\\
				\end{bnfsplit} }\\
			\bnfprod{MDEF}
			{\dots \bnfor
				\bnfpn{P},\ \bnfpn{M} = \texttt{mut\_mkbor\_mem2reg} \bnfsp \bnfpn{P},\ \bnfsp \bnfpn{M} \bnfor \bnfpn{M} = \texttt{mov\_reg2mem} \bnfsp \bnfpn{P}, \bnfsp \bnfpn{P}, \bnfsp \bnfpn{M}} \\
			\bnfprod{OS}
			{\texttt{die} \bnfsp \bnfpn{P}}
		\end{bnf*}
	\end{minipage}
	\caption{Ownership instr. in \oseair grammar, where \texttt{R}, \texttt{P}, and \texttt{M} are scalar registers, pointer registers, and memory registers respectively.}
	\label{fig:oseair-syntax}
  \end{figure*}

\bparagraph{Syntax.}
We introduce ownership semantics on the base language \seair~\cite{DBLP:conf/fmcad/PriyaSBZVG22}.
\seair is an intermediate representation (IR) itself based on \llvm IR.
\llvm assumes a register based machine and dependency between memory operations are implied.
\seair explicates this dependency information between memory operations by introducing memory registers.
We assume that the type of each register is known.
\Cref{fig:oseair-syntax} shows the ownership extended syntax of \seair called \oseair.
We use \code{R} to represent a scalar register, \code{P} for a pointer register and \code{M} for a memory register.
A legal \oseair program is assumed to be in a Static Single Assignment (SSA) form.
\oseair primarily replaces unrestricted alias creation by new operations that introduce and remove aliases in a restricted manner.
The \mkown instruction initializes memory at the given location (simlar to a \code{Box::new(n)} in \rust).
The \code{mut_mkbor}, \code{mut_mksuc} instructions occur in pairs.
The first creates a mutable borrow pointer from a lender pointer.
The second creates a succeeding pointer from the lender pointer that becomes active after the mutable borrow ends.
The \code{mut_mkbor_off} is similar to a \code{mut_mkbor} and creates a pointer at an offset within an allocation.
It must be followed by a \code{mut_mksuc} instruction.
The \code{ro_*} instructions create read-only borrows of the lending pointer.
The \code{cpy_*} instructions create unrestricted copies of the lending pointer.
The \code{mut_mkbor_mem2reg} instruction borrows (loads) a pointer stored in memory to a register.
The \code{mov_reg2mem} instruction moves (stores) a pointer in a register to memory.
There is no move instruction between registers since the operation is equivalent to $\alpha$-renaming.

\begin{table*}[t]
	\centering
	\renewcommand{\arraystretch}{1.3} 
	\begin{tabular}{l|l|l}
		\bf Operation                                                                                                                                                                                                   & \bf Pre-condition             & \bf Post-condition                           \\
		\hline
		\code{p,m1 = mkown n, m0}                                                                                                                                                                                       & $S_{B}[p.\loc] = \varnothing$ & $S_{B}[p.\loc] = (tag_{p},\mathsf{o}) :: []$ \\
		\hline
		\begin{tabular}{@{}l@{}}
			\code{q0 = mut\_mkbor p0} \\
			\code{p1 = mut\_mksuc p0}
		\end{tabular}                                                                                                                                                                                    &
		\begin{tabular}{@{}l@{}}
			$S_{B}[p_{0}.\loc] = B_{0} :: (tag_{p_{0}}, t) :: B_{1},$ \\
			$t \in \{\mathsf{o}, \mathsf{mb} \}, p_{0}.tag = tag_{p_{0}}$
		\end{tabular}                                                                     &
		$S_{B}[p_{0}.\loc] = (tag_{q_{0}}, \mathsf{mb}) :: (tag_{p_{1}}, t) :: B_{1}$                                                                                                                                                                                                                  \\
		\hline
		\begin{tabular}{@{}l@{}}
			\code{c1 = cpy\_mkcpy1 p0} \\
			\code{c2 = cpy\_mkcpy2 p0}
		\end{tabular}                                                                                                                                                                                   &
		\begin{tabular}{@{}l@{}}
			$S_{B}[p_{0}.\loc] = B_{0} :: (tag_{p_{0}}, t) :: B_{1},$ \\
			$p_{0}.tag = tag_{p_{0}}$
		\end{tabular}                                                                  &
		$S_{B}[p_{0}.\loc] = (tag_{c_{1}}, \mathsf{c}) :: (tag_{c_{2}}, t) :: B_{1}$                                                                                                                                                                                                                   \\
		\hline
		\code{die q}                                                                                                                                                                                                    &
		\begin{tabular}{@{}l@{}}
			$S_{B}[q.\loc] = (tag_{q}, t_{q}) :: (tag_{p}, t_{p}) ::  B_{1}, $ \\
			$q.tag = tag_{q}, t_{q} = \mathsf{mb},
				t_{p} \in \{\mathsf{o},\mathsf{mb}\}$
		\end{tabular} &
		$S_{B}[q.\loc] = (tag_{p}, t_{p}) :: B_{1}$                                                                                                                                                                                                                                                    \\
		\hline
		\code{m1 = store r, p, m0}                                                                                                                                                                                      &
		\begin{tabular}{@{}l@{}}
			$S_{B}[p.\loc] = B_{0} :: (tag_{p},t_{p}) :: B_{1},$ \\
			$t_{p} \ne \mathsf{rb}, p.tag = tag_p$
		\end{tabular}                                                               &
		\begin{tabular}{@{}l@{}}
			$S_{B}[p.\loc] = B_{2} :: (tag_{p},t_{p}) :: B_{1},$                                \\
			$(p.tag = tag_p, t_p \in \{\mathsf{o},\mathsf{mb}\}) \implies (B_2 = \varnothing),$ \\
			$(t_p=c) \implies (B_2=B_0)$
		\end{tabular}                                                                                                                  \\
		\hline
		\code{r = load p, m}                                                                                                                                                                                            &
		\begin{tabular}{@{}l@{}}
			$S_{B}[p.\loc] = B_{0} :: (tag_{p},t_{p}) :: B_{1},$ \\
			$p.tag = tag_{p}$                                    \\
		\end{tabular}                                                               &
		\begin{tabular}{@{}l@{}}
			$S_{B}[p.\loc] = B_{2} :: (tag_{p},t_{p}) :: B_{1},$                                                              \\
			$(t_p \in \{\mathsf{o},\mathsf{mb},\mathsf{rb}\}) \implies (B_2 = [(tag_q, t_q) \in B_0 \mid t_{q}=\mathsf{c}]),$ \\
			$(t_p=c) \implies (B_2=B_0)$
		\end{tabular}
	\end{tabular}
	\caption{Effect of selected operations on borrow stack ($S_{B}$) in machine \machzero. Effects on $R$ and $M$ are not shown.}
	\label{tab:effect_borrow_stack}
	\vspace{-0.25in}
\end{table*}


\bparagraph{Semantics of \machzero.}
The semantics are given in terms of a machine \machzero and is based on the stacked borrows model for \rust~\cite{DBLP:journals/pacmpl/JungDKD20}.
In our formulation, each pointer type ($ptype$) is one of owned ($\mathsf{o}$), mutably borrowed ($\mathsf{mb}$), immutably borrowed ($\mathsf{rb}$), or, copied ($\mathsf{c}$).
Access to memory is controlled by maintaining a per-location borrow stack that captures both valid accessors and access order.
The configuration of \machzero is given by the program counter state ($P$), register map ($R : \text{id} \rightarrow value$), address map ($M : \loc \rightarrow value$) and a borrow store state ($S_{B}:\loc \rightarrow stack((tag, ptype))$).
A $value$ is either a bit-vector $bv(N)$ or a pointer type.
An address ($\loc$) is represented as a bit-vector.
A pointer is a tuple of $(\loc,tag)$ and is considered \emph{fat} on account of additional metadata carried along with the address\footnote{We use the shorthand $.\loc$ to refer to the first tuple element, similarly for other elements.}.
A $tag:bv(N)$ is a unique id given to a pointer when it is defined.
Operations that introduce and remove aliases, then push and pop alias tags on the borrow stack, respectively.
Each borrow stack entry also stores $ptype$ along with an identifier for finer access control.
An important restriction is that memory access is allowed for an alias if its $tag$ is top-of-(borrow)stack for that address.

The semantics for relevant pointer introduction, aliasing, and, removal are given through operations on the borrow stack ($B$) in~\cref{tab:effect_borrow_stack}.
A borrow stack state is represented as a list $ B = e :: B_{1}$, where $e$ is the top of stack and $B_{1}$ represents the rest of the stack.
We do not explicitly show effect of operations on $(P,R,M)$ nor do we give the semantics for all instructions of \machzero due to space constraints.
The interested reader is referred to the stacked borrows~\cite{DBLP:journals/pacmpl/JungDKD20} and \seair~\cite{DBLP:conf/fmcad/PriyaSBZVG22} papers for further background.
The \mkown operation allocates and stores $n$ at the given location.
This operation must provide a location that is un-allocated.
After the operation, the new pointer $tag$ is pushed onto the stack with $ptype=\mathsf{o}	$.
The mutable borrow operations use \mutborbor, \mutborsuc instructions that always occur in a pair on a lender pointer $p_{0}$ to create a borrowed pointer $q_{0}$ and a succeeding pointer $p_{1}$.
For a successful operation, the borrow stack is popped until $p_{0}$ is on top and its $ptype$ is either an owning or a mutably borrowed pointer.
This operation removes $p_{0}.tag$ as an accessor and instead pushes $p_{1}.tag$, the succeeding pointer and $q_{0}.tag$, the borrowed pointer, to the borrow stack in that order.
The associated type of a pointer is also added to each stack entry.
Note that the type of  $q_{1}$ is always $\mathsf{mb}$.
However the type of $p_{1}$ depends on the type of $p_{0}$.
The intent is for $q_{0}$ to have access rights till it surrenders them to $p_{1}$.

The copy operation creates two copies $c_{1}$ and $c_{2}$ using \cpyone and \cpytwo instructions.
A copied pointer corresponds to a raw pointer in \rust.
The lender pointer $p_{0}$  for a copy operation can be of any $ptype$.
Similar to a mutable borrow operation, all entries on top of $p_{0}$  are popped from the borrow stack and $p_{0}$ itself is removed.
Next $c_{1}::c_{2}$ are pushed onto the borrow stack in that order.
The $ptype$ of $c_{1}$ is always $c$.
However, the $ptype$ of $c_{2}$ depends on the lender pointer $p_{0}$.
This ensures that the $ptype$ of a lender pointer is not lost through successive copy operations.
Finally, the \die operation surrenders access rights for a pointer by popping off its entry from the borrow stack.
It is only defined for a mutably borrowed pointer $q$ and signals transfer of data from such a pointer to its immediate lender, which must be of $ptype=\mathsf{o}$ or $ptype=\mathsf{mb}$.
The pointer $q$ must be top of borrow stack.
The \die operation is an extension of stacked borrows and is useful for returning information from a mutable borrow to the succeeding pointer without going through shared memory.
The \code{store} instruction writes a value to memory.
If the lender pointer $p$ is mutably borrowed or owning then all elements before $p$ are popped.
If $p$ is copied then borrow stack remains unchanged.
The \code{load} instruction reads values from memory into a register using a lender pointer $p$.
If $p$ is owning, mutably borrowed or read-only borrowed, then all pointers above $p$ (except copied pointers) are removed from the borrow stack.
If $p$ is copied, then the borrow stack is unchanged.
Finally, the observable state $ObsState_{\machzero}$ of machine \machzero is given by the tuple $(P,R,M,S_{B})$.

Let us look at an example of how \machzero operates in~\cref{fig:m0_example}.
The intent of the program is to
\begin{inparaenum}[(1)]
	\item create an owned pointer,
	\item make its alias
	\item update data through the alias, and,
	\item observe the data through the owned pointer.
\end{inparaenum}
At line~\ref{line:mkown-m0-ex}, a word of memory is allocated with \code{(addr=0x4,tag=1)} in the register map at key \code{p0}, the integer \code{42} is written to memory at \code{M[0x4]}, and the $tag$ value $1$ is pushed to the borrow stack at \code{SB[0x4]}.
Next an alias is created using the mutable borrow operation at lines~\ref{line:mkbor-m0-ex}--\ref{line:mksuc-m0-ex} using tags \code{3} and \code{2} for borrowed \code{q0} and succeeding pointer \code{p1} respectively.
First the tag for \code{p1} is pushed, then the tag for \code{q0} is pushed.
The next couple of lines load \code{42} using \code{q0}, increment it, and write it back.
The program ends the mutable borrow in line~\ref{line:die-m0-ex}.
This removes \code{q0}'s tag from \code{SB}.
Now only \code{p1} can access $\loc$ \code{0x4}.
Finally, the program reads the new value \code{43} from $\loc$ \code{0x4} in line~\ref{line:load-m0-ex}.

\newsavebox{\figopsemex}
\begin{lrbox}{\figopsemex}%
	\begin{lstlisting}[language=seair]
fun main() {
BB0:
  m00 = mem.init()
  ;*@
	\begin{tabular}{l|l|l}
	  \cellcolor{pink!60} R = [] & \cellcolor{pink!60} M = [] & \cellcolor{pink!60} SB = []
	\end{tabular}
  @*
  p0,m0 = mk_own 42, m00*@\label{line:mkown-m0-ex}@*
  ;*@
	\begin{tabular}{c|c|c}
		\cellcolor{pink!60} R[p0] = (0x4,1) & \cellcolor{pink!60} M[0x4] = 42 & \cellcolor{pink!60} SB[0x4] = 1 :: []
	\end{tabular}
  @*
  q0 = mut_mkbor p0*@\label{line:mkbor-m0-ex}@*
  p1 = mut_mksuc p0*@\label{line:mksuc-m0-ex}@*
  ;*@
	\begin{tabular}{c|c|c}
	  \cellcolor{pink!60} R[p1] = (0x4,2) & \cellcolor{pink!60} & \cellcolor{pink!60} \\
	  \cellcolor{pink!60} R[q0] = (0x4,3) & \cellcolor{pink!60} M & \cellcolor{pink!60} SB[0x4] = 3 :: 2 :: []
	\end{tabular}
  @*
  r1 = load q0, m0
  *@
	\begin{tabular}{c|c|c}
	  \cellcolor{pink!60} R[r1] = 42 & \cellcolor{pink!60} M & \cellcolor{pink!60} SB
	\end{tabular}
  ;@*
  m1 = store r1 + 1,q0,m0
  ;*@
	\begin{tabular}{c|c|c}
	  \cellcolor{pink!60} R & \cellcolor{pink!60} M[0x4] = 43 & \cellcolor{pink!60} SB
	\end{tabular}
  @*
  die q0*@\label{line:die-m0-ex}@*
  ;*@
	\begin{tabular}{c|c|c}
	  \cellcolor{pink!60} R & \cellcolor{pink!60} M & \cellcolor{pink!60} SB[0x4] = 2 :: []
	\end{tabular}
  @*
  r = load p1, m1*@\label{line:load-m0-ex}@*
  ;*@
	\begin{tabular}{c|c|c}
	  \cellcolor{pink!60} R[r] = 43 & \cellcolor{pink!60} M & \cellcolor{pink!60} SB
	\end{tabular}
  @*
  halt
}
\end{lstlisting}
\end{lrbox}%
\begin{figure}[t]
	\centering
	\scalebox{1.0}{\usebox{\figopsemex}}
	\caption{Example of \machzero operation. Effect on register map ($R$), memory map ($M$), and borrow store ($S_{B}$) shown in \marktext[pink!60]{pink}.}
	\label{fig:m0_example}
	\vspace{-0.25in}
\end{figure}

\bparagraph{Semantics of \machone.}
We now define an extension to \machzero called \machone.
In \machone, a fat pointer additionally has a \emph{cache} bit-vector field called $\val$.
Each store operation also updates $\val$ with the value to be written to memory.
A load from memory may be replaced by $\val$ when correct to do so.
A pointer value now becomes $(\loc,tag,\val)$.
Overall, the semantics of existing instructions aim to maintain the $\val$ cache.
The semantics is laid out in~\cref{tab:effect_m1}.
\begin{table*}[t]
	\centering
	\renewcommand{\arraystretch}{1.3} 
	\begin{tabular}{l|l|l}
		Operation                                                                                                         & Pre-condition & Post-condition                                            \\
		\hline
		\code{p = mkown n}                                                                                                & --            & $R[\mathsf{p}]=(p.\loc,tag_{p},n),M[p.\loc]=n$            \\
		\hline
		\begin{tabular}{@{}l@{}}
			\code{q0 = mut\_mkbor p0} \\
			\code{p1 = mut\_mksuc p0}
		\end{tabular}                                                                                      &
		\begin{tabular}{@{}l@{}}
			$R[\mathsf{p_{0}}]=(p_{0}.\loc,tag_{p_{0}},v_{p})$
		\end{tabular} &
		\begin{tabular}{@{}l@{}}
			$R[\mathsf{q_0}]=(p_{0}.\loc,tag_{q_{0}},v),R[\mathsf{p_1}]=(p_{0}.\loc,tag_{p_1},v),$ \\$
			(B_{0} = \varnothing) \implies ( v =v_p),$                                           \\$
				(B_{0} \neq \varnothing) \implies (v=M[p_{0}.\loc])$
		\end{tabular} \\
		\hline
		\begin{tabular}{@{}l@{}}
			\code{c1 = cpy\_mkcpy1 p0} \\
		  \code{c2 = cpy\_mkcpy2 p0}
		\end{tabular}                                                                                     &
		\begin{tabular}{@{}l@{}}
			$R[\mathsf{p_{0}}]=(p_{0}.\loc,tag_{p_0},v_{p})$
		\end{tabular}                            &
		\begin{tabular}{@{}l@{}}
			$R[\mathsf{c_1}]=(p_{0}.\loc,tag_{c_1},v),R[\mathsf{c_2}]=(p_{0}.\loc,tag_{c_{2}},v),$ \\$
			(B_{0} = \varnothing \land{} t = \{\mathsf{o}, \mathsf{mb}, \mathsf{rb}\}) \implies ( v=v_p),$                                              \\$
					\neg (B_{0} = \varnothing \land{} t = \{\mathsf{o}, \mathsf{mb}, \mathsf{rb}\}) \implies (v=M[p_{0}.\loc])$
		\end{tabular} \\
		\hline
		\code{die q}                                                                                                      &
		\begin{tabular}{@{}l@{}}
			$R[\mathsf{q}]=(q.\loc,tag_q,n)$
		\end{tabular}                                                                 &
		\begin{tabular}{@{}l@{}}
			$R[p]=(q.\loc,tag_{p},n),$ \\$\exists p.R[p]=(q.\loc,tag_p,\_)$
		\end{tabular}                                                                                                      \\
		\hline
		\code{m1 = store r, p, m0}                                                                                        &
		\begin{tabular}{@{}l@{}}
			$R[\mathsf{p}]=(p.\loc,tag_p,\_)$
		\end{tabular}                                                                &
		\begin{tabular}{@{}l@{}}
			$M[p.\loc] = v,R[\mathsf{p}]=(p.\loc,tag_{p},v)$
		\end{tabular}                                                                                                              \\
		\hline
		\code{r = load p, m}                                                                                              &
		\begin{tabular}{@{}l@{}}
			$R[\mathsf{p}]=(p.\loc, tag_p, v_{p})$
		\end{tabular}                                  &
		\begin{tabular}{@{}l@{}}
			$R[\mathsf{r}]=v,R[\mathsf{p}]=(p.\loc, tag_p, v),$                             \\
			$((B_0 = \varnothing, t_{p} \in\{\mathsf{o},\mathsf{mb}\}) \implies (v = v_p))$ \\
			$((B_0 \ne \varnothing \lor{} t_{p}=\mathsf{c}) \implies (v = M[p.\loc]))$
		\end{tabular}                                                           \\
	\end{tabular}
	\caption{Effect of selected operations on $S_{B}$,$R$, and $M$ in machine \machone in addition to pre-and-post conditions from~\cref{tab:effect_borrow_stack}.}
	\label{tab:effect_m1}
	\vspace{-0.25in}
\end{table*}
The \mkown instruction updates its cache with the value it initialized the memory allocation with.
The pair of \mutborbor and \mutborsuc operations have two cases:
\begin{inparaenum}[(1)]
	\item if the lender is top-of-(borrow)stack then the operation reads the value stored at lender pointer $p_{0}$ and updates the caches of $q_{0}$ and $p_{1}$ with that value;
	\item if the lender is not top of stack then the value at lender may be stale and the correct value is read from memory.
\end{inparaenum}
The pair of \cpyone and \cpytwo instructions similarly update the cache of $c_{1}$ and $c_{2}$ with the correct value.
The \die instruction transfers the value cached at $q$ to the cache of the immediately succeeding pointer, called $p$ here.
The transfer to the succeeding pointer occurs by first searching for the pointer with the correct $tag$ in the register map $R$ and then updating the corresponding $\val$ field.
Since we do not support the storage of pointers to memory, the search through $R$ is enough to find the right pointer.
Note that the \code{die} operation enables transfer of a value from a mutable borrow to the succeeding pointer without using shared memory.
A \code{store} instruction updates the cache with the value \code{r} to be written to memory.
This value is then written to memory and to $p.\val$.
In \machone, a \code{store} does not support storing pointers to memory.
This restriction is lifted in
\ifarxivmode
	~\cref{app:appendixa}.
\else
	an extended version~\cite{priya2024ownershiplowlevelintermediaterepresentation}.
\fi
A \code{load} has two cases.
First, if the lender pointer $p$ is top-of-(borrow)stack, and is mutably borrowed or owning, then the read from memory is replaced by a read of the $\val$ (cache) field.
Second, if the \code{load} uses a lender pointer $p$ that is not top-of-(borrow)stack, or is copied, then the read from memory proceeds as usual.
In the second case, the pointer cache is also updated with the value read from memory.

The optimisation we describe for the \code{load} instruction is correct because \machone always maintains the following invariant:
\begin{theorem}[Cache equivalence]
	\label{thm:cache-equiv}
	For all pointers in the register map, if the pointer is top-of-(borrow)stack and is owning or mutably borrowed then the pointer cache value is the same as the value of memory at address of the pointer. Formally, let $R$ be a register map, $M$ memory, and $S_B$ a borrow store. Then,
	\begin{multline*}
		(R[p] = (\loc, tag_{p}, n)) \land{}\\ (S_{B}[\loc] = (tag_{p},t_{p}) :: B) \land{}\\ (t_{p} \in \{\mathsf{o},\mathsf{mb}\})) \implies M[\loc] = n
	\end{multline*}
\end{theorem}
\gdef\proofSymbol{\ensuremath{\blacksquare}}
\begin{proof}
	The proof proceeds by  structural induction on the syntax of the program P.
	Assume Thm.~\ref{thm:cache-equiv} holds in some configuration $(P_{0},R_{0},M_{0},S_{B_{0}})$.
	The next instruction takes the configuration to $(P_{1},R_{1},M_{1},S_{B_{1}})$.
	We case-split on each possible instruction. We illustrate the process through some of the relevant instructions.
	\begin{itemize}
		\item \code{store} keeps the cache in-sync with memory according to given semantics;
		\item \code{mut_mkbor} keeps the mutably borrowed pointer cache in-sync with memory since the lender cache value is already in-sync (by assumption) and mutably borrowed pointer cache gets this value;
		\item \code{die}, before this \code{die} Thm.~\ref{thm:cache-equiv} holds for the mutably borrowed pointer. Then, \code{die} copies cache value from mutably borrowed pointer to succeeding pointer, keeping the succeeding pointer cache in-sync with memory.
	\end{itemize}
\end{proof}

We now define $ObsState_{M1}$ for \machone as a tuple $(P,R,M,S)$ with the pointer $\val$ field excluded from view.
Let $\equiv$ be the equivalence relation between \machzero and \machone defined as follows: $s_{M0}^{m_{0}} \equiv s_{M1}^{m_{1}} \;\leftrightarrow\; ObsState_{M0}(s_{M0}) = ObsState_{M1}(s_{M1})$.
By Thm.~\ref{thm:cache-equiv}, starting in equivalent observable states, both \machzero and \machone operate in lock-step. Thus, the following theorem holds:
\begin{theorem}
	\label{thm:equiv}
	The relation $\equiv$ is both a forward and a backward simulation between \machzero and \machone.
\end{theorem}
Thus, safety of \machone implies safety of \machzero and vise versa.









\section{VC Generation}
\label{sec:vcgen}
\newsavebox{\figpropex}
\begin{lrbox}{\figpropex}%
	\begin{lstlisting}
fun main() {
BB0:
  m00 = mem.init()
  ;*@\label{line:meminit-prop-ex}@**@\marktext{$m_{00}$}@*
  p0,m0 = mk_own 42, m00
  ;*@\label{line:mkown-prop-ex}@**@\marktext{$p_{0}.\loc=4 \land{} p_{0}.\val=42 \land{}$}@*
  ;*@\marktext{$m_{0} = m_{00}[p_{0}.\loc \mapsto 42]$}@*
  q0 = mut_mkbor p0
  p1 = mut_mksuc p0
  ;*@\label{line:mkbor-prop-ex}@**@\marktext{$q_{0}.\loc=p_{0}.\loc \land{} q_{0}.\val=p_{0}.\val \land{} q_{0}.\ret\val=x \land{}$}@*
  ;*@\label{line:mksuc-prop-ex}@**@\marktext{$p_{1}.\loc=p_{0}.\loc \land{} p_{1}.\val=x \land{} p_{1}.\ret\val=p_{0}.\ret\val$}@*
  r1 = load q0, m0
  ;*@\label{line:load-prop-ex}@**@
  \marktext{$r_{1}=q_{0}.\val$}
  @*
  m1 = store r1 + 1, q0, m0
  ;*@\label{line:store1-prop-ex}@**@
  \marktext{$q_{1}.\loc = q_{0}.\loc \land{} q_{1}.\ret\val = q_{0}.\ret\val \land{}$}
  @*
  ;*@\label{line:store2-prop-ex}@**@
  \marktext{$q_{1}.\val = r_{1} + 1 \land{} m_{1} = m_{0}[q_{1}.\loc \mapsto q_{1}.\val]$}
  @*
  die q0
  ;*@\label{line:die-prop-ex}@**@
  \marktext{$q_{1}.\val = q_{1}.\ret\val$}
  @*
  r = load p1, m1*@\label{line:load-m0-ex}@*
  ;*@
  \marktext{$r=p_{1}.\val$}
  @*
  assert r == 43
  ;*@
  \marktext{$\neg (r = 43)$}
  @*
  halt
}
\end{lstlisting}
\end{lrbox}%
\begin{figure}[t]
	\centering
	\scalebox{1.0}{\usebox{\figpropex}}
	\caption{Verification condition (VC) shown in \marktext{yellow}.}
	\label{fig:prop-example}
	\vspace{-0.25in}
\end{figure}
We introduce the general encoding of an \oseair program and the modelling of mutable borrows in particular using the example in~\cref{fig:prop-example}.
Note that this example runs throughout this section.
For now, we suggest the reader ignore the generated VC (in yellow).
We focus on aliasing instructions and how the pointer cache is affected.
The \code{mk_own} instruction defines \code{p0}  writing \code{42} to both memory and the pointer cache maintaining \emph{Cache Equivalence}.
The \code{mut_mkbor}, \code{mut_mksuc} instructions create aliases \code{q0}, \code{p1} from \code{p0}.
Here, the cache at \code{p0} is copied to \code{q0} and \code{p1}, again maintaining the cache equivalence invariant.
The \code{q0} mutably borrowed alias updates memory (and its pointer cache) to \code{43}.
It then surrenders access rights using the \code{die} instruction.
At this point, the succeeding alias \code{p1} becomes active (top-of-(borrow)stack).
However, for \code{p1} to maintain cache equivalence (Theorem \ref{thm:cache-equiv}), it must get a copy of \code{q0}'s cache.
This is not straightforward since there is no explicit transfer instruction from \code{q0} to \code{p1}.
The standard solution is to use shared memory so that \code{q0} can write to this memory on a \code{die} and the succeeding pointer \code{p1} can then read from this memory on next access.
However, the aim of caching is to eschew memory accesses as much as possible to keep the operation (and VC) simple.
The concrete semantics of \machone provides one alternative to accessing memory.
There, a \code{die} instruction finds the succeeding pointer $tag$ in the borrow store $S_{B}$ and then searches through the register map $R$ to update the pointer cache with the same $tag$.
This mechanism is as (or more) expensive to execute symbolically as shared memory.
An elegant solution proposed in RustHorn~\cite{DBLP:journals/toplas/MatsushitaTK21} uses a \emph{prophecy variable}~\cite{DBLP:journals/tcs/AbadiL91} to model the return of a mutable borrow in the VC.
We adapt the scheme to VC generation (VCGen) for \oseair.
We now explain VCGen, emphasizing the role of prophecy variables to model return of a mutable borrow.

The VC is generated using the \symfun translation function.
It builds up the VC in a recursive, bottom-up fashion on the abstract syntax tree of an \oseair program.
For simplicity of presentation, we assume that two fundamental sorts are used in the encoding: bit-vector of $64$ bits, \bv{64}, and a map between bit-vectors, $\bv{64} \to \bv{64}$.
We now revisit the example and explain the VC for each line of source code.
Line~\ref{line:meminit-prop-ex} models \code{mem.init} as $m_{00}$,  a free variable.
Line~\ref{line:mkown-prop-ex} models the \mkown instruction.
It updates memory at  $m_{00}[\loc]$ to $42$ and defines the fat pointer $p_{0}$.
A fat pointer is modelled as a tuple $(\loc,\val,\ret\val)$.
Here $\loc$ holds the address, $\val$ holds the current cache value ($42$ here), and $\ret\val$ holds a prophecy value, the use of which will be laid out soon.
A mutable borrow operation occurs in lines~\ref{line:mkbor-prop-ex}--\ref{line:mksuc-prop-ex}.
The lender pointer $p_{0}$ creates two aliases, the mutable borrow $q_{0}$ and the succeeding pointer $p_{1}$. The location
$p_{0}.\loc$ is copied to both $q_{0}.\loc$ and $p_{1}.\loc$.
The cache at $p_{0}.\val$ is copied to $q_{0}.\val$.
To set up the return of the cache value from the mutably borrowed alias to the succeeding pointer, we \emph{entangle} the $q_{0}.\ret\val$ and $p_{1}.\val$ field using a fresh prophecy value $x$.
This prophecy $x$ will resolve to the correct cache value when \code{q0} dies.
When this happens, $p_{1}$ instantly gets the same value in its cache in $p_{1}.\val$.
Moving ahead, lines~\ref{line:load-prop-ex}--\ref{line:store2-prop-ex} model the increment of the value pointed to by $q_{0}$.
Note that apart from updating the value in memory, the $q_{0}.\val$ variant $q_{1}.\val$ also gets the updated value.
Finally, in line~\ref{line:die-prop-ex}, the \die operation causes the prophecy $x$ to be constrained by equating $q_{1}.\val$ and $q_{1}.\ret\val$.
As expected, this defines $p_{1}.\val$ to get the correct cache value $43$ maintaining cache equivalence.
The transfer of cache from \code{q0} to \code{p1} is, therefore, modelled without any expensive symbolic operations involving memory accesses or register map lookups.
In the end, we see that the generated VC is unsatisfiable and the property is valid.

We now describe the function \symfun for selected pointer operations.
The semantics of~\code{mk_own} is given in~\cref{fig:own_sem_ptr_create}.
We assume that an address $\ell$ is given by an external allocator.
The allocator should follow the usual property that $\ell$ has not been allocated previously.
Note that $p_{0}.\ret\val$ field is free since an owning pointer does not return the cache value to another alias.
We define \symfun for mutable borrow and die operations in~\cref{fig:own_sem_ptr_borrow}.
The mutable borrow aliasing operation copies the \loc field from the lender to the borrower and succeeding pointer.
The cache is wired as follows.
First, the mutably borrowing pointer gets the lender cache using $q_{0}.\val=p_{0}.\val$.
Second, we entangle $p_{1}.\val$ with the free symbol $q_{0}.\ret\val$ using the \entangle macro.
The macro itself entangles the first argument with the second by equating them.
Third, $p_{1}.\ret\val$ gets the prophecy in $p_{0}.\ret\val$ to model cascading borrows (reborrows).
The \symfun for \emph{die} equates the given pointer's $\val$ and $\ret\val$ field, constraining the prophecy value in $q.\ret\val$ and returning the borrow.

In summary, the fat pointer concept is our workhorse in mapping two previous high level VCGen schemes to a low-level verification setting.
First, the reference elimination mechanism is replaced by fat pointers that cache values.
Second, a fat pointer field holds a prophecy value that expresses the cache value after returning from a mutable borrow.
\begin{figure}
	\scriptsize
	\begin{multline*}
		\sym{\tcode{p0, m1 = mk\_own n, m0}} \sdef
		\exists \ell . (m_{1} = m_{0}[\ell \mapsto n]) \land{} \\
		(p_{0}.\loc = \ell) \land{} (p_{0}.\val = n)
	\end{multline*}
	\caption[Own Semantics (v1)]{Definition of \symfun for \code{mk_own}.}
	\label{fig:own_sem_ptr_create}
	\vspace{-0.25in}
\end{figure}

\begin{figure}
	\scriptsize	\begin{align*}
     & \entangle(r_{1}, r_{2}) \sdef{} r_{1} = r_{2} \\
    & \sym{\tcode{q_{0} = mut\_mkbor p_{0}};\ \tcode{p_{1} = mut\_mksuc p_{0}}} \sdef{}                                                \\&\quad\quad
                                                                                                                                                                                                                                                                                                                                                                                                                                                                  q_{0}.\loc = p_{0}.\loc \land{}                                                                                                          q_{0}.\val = p_{0}.\val \land{} \\&\quad\quad\quad\quad
                                                                                                                                                                                                                                                                                                                                                                                                                                                                 \entangle(p_{1}.\val, q_{0}.\ret{\val}) \land{}
                                                                                                                                                                                                                                                                                                                                                                                                                                                                  p_{1}.\ret{\val} = p_{0}.\ret{\val}                                                                                                 \\
                                                                                                                           & \sym{\tcode{die q}} \sdef{} q.\val = q.\ret{\val}
	\end{align*}
	\caption[Own Semantics (v1)]{Definition of \symfun for mutable borrow, die, and \entangle macro for entanglement.}
	\label{fig:own_sem_ptr_borrow}
	\vspace{-0.25in}
\end{figure}

\section{Towards a practical machine}
\label{sec:multiword}
\begin{figure}[t]
	\centering
	\begin{minipage}{0.9\linewidth}
		\scriptsize
		\begin{bnf*}
			\bnfprod{RDEF}{
				\begin{bnfsplit}
					\dots
					\bnfor
					\bnfpn{P}, \bnfpn{M} = \texttt{mk\_own} \bnfsp \bnfpn{R}, \bnfpn{M}
					\bnfor\\
					\bnfpn{P} = \texttt{begin\_unique} \bnfsp \bnfpn{P}
					\bnfor
					\bnfpn{P} = \texttt{end\_unique} \bnfsp \bnfpn{P}
					\bnfor
					\\
					\bnfpn{P} = \texttt{set\_cache} \bnfsp \bnfpn{P}, \bnfsp \bnfpn{R}
					\bnfor
					\bnfpn{R} = \texttt{get\_cache} \bnfsp \bnfpn{P}
				\end{bnfsplit} }
		\end{bnf*}
	\end{minipage}
	\caption{Grammar of new instructions for \oseair.}
	\label{fig:oseair-multi-syntax}
	\vspace{-0.25in}
  \end{figure}
In ~\cref{sec:opsem}, we described \machzero and \machone, both machines that could only allocate a single word through \code{mk_own}.
We lift this restriction now in \machtwo.
To allocate multiple words (wide allocations), we change the \code{mk_own} syntax.
Instead of taking a bit-vector to write to memory, it now takes a bit-vector \emph{allocation size} argument.
For cache equivalence to hold, the pointer cache width must now be wide enough to cache multi-byte allocation data.
This complicates the design of the cache.
To keep things simple,
instead of hard-wiring the pointer cache to replicate memory contents, we only cache a \emph{summary} of the data in memory and provide operations to set and get the cache value using \setcache and \getcache, respectively.
A property to be verified can be cached at the pointer. Pointer aliasing operations copy the value as before.
The decoupling of cache from \code{load} and \code{store} operations does introduce burden on the programmer to update the cache as required.
As we move towards a practical machine, we also add a new unique ($\mathsf{u}$) variant to pointer type $ptype$.
A unique pointer is created using \beginunique and \uniqueend instructions.

The syntax of these new instructions is given in~\cref{fig:oseair-multi-syntax}.
The \code{mk_own} instruction takes three arguments - the bit-vector to write, the size (in bytes) of the allocation and the incoming memory to update.
The operation now does not update memory or the pointer cache since that is the programmer's responsibility.
The \beginunique and \uniqueend operations take a copied (unique) pointer and define a unique (copied) pointer with the same $\loc$ and $\val$ fields as the source pointer.
These operations are useful in practical situations where the user only wants to mark a pointer as unique temporarily.
The \code{get_cache} instruction returns the $\val$ field of a pointer.
the \code{set_cache} instruction takes a pointer and a value.
It then defines a new pointer where all fields are the same as the source pointer, except the $\val$ field that has been updated to the given value.
\begin{figure}[t]
  \centering
	\begin{lstlisting}[language=C]
extern void escapeToMemory(char *);
int main() {
  *@\marktext{char *p = MK\_OWN(0, sizeof(char));}@* *@\label{line:own-example-C-mkown} @*
  char c = nd_char();
  assume (c == 42);
  *@\marktext{SET\_CACHE(p, c);}@* *@\label{line:own-example-C-setcache} @*
  *p = c;
  char *b;
  *@\marktext{MUT\_BORROW(b, p);}@**@\label{line:own-example-borrow}@*
  if (nd_bool()) {*@\label{line:own-example-cond}@*
    c = nd_char();
    assume(c > 43);
    *@\marktext{SET\_CACHE(b, c);}@*
    *b = c;
    escapeToMemory(b);
  }
  *@\marktext{DIE(b);}@**@\label{line:own-example-die} @*
  char r;
  *@\marktext{GET\_CACHE(p, r);}@**@\label{line:own-example-getcache} @*
  sassert(r == 42 || r > 43);
  return 0;}
\end{lstlisting}%
	\caption{An example C program with ownership intrinsics wrapped in C Macros, highlighted in \marktext{yellow}.}
	\label{fig:own-example-C}
	\vspace{-0.25in}
\end{figure}%

\newsavebox{\figownrpd}
\begin{lrbox}{\figownrpd}%
	\begin{lstlisting}
fun main() {
BB0:
  m3 = mem.init()
  *@\marktext{p2, m0 = mk\_own 1, m3}@*
  r5 = nd_char()
  r6 = r5 == 42
  *@\marktext{p3 = set\_cache p2 r5}@**@\label{line:own-oseair-setcache}@*
  m1 = store r5, p3, m0
  *@\marktext{p5 = mut\_mkbor p3}@**@\label{line:own-oseair-mkbor}@*
  *@\marktext{p6 = mut\_mksuc p3}@**@\label{line:own-oseair-mksuc}@*
  r15 = nd_bool();
  r17 = r15 == 42
  br r17, ERR, BB1

BB1:
  r18 = nd_char()
  r19 = r18 > 43
  r20 = r6 && r19
  *@\marktext{p23 = set\_cache p5 r18}@**@\label{line:own-oseair-setcache2}@*
  m2 = store r18, p23, m1 *@\label{line:own-oseair-store}@*
  escapeToMemory(p0)
  br ERR

ERR:
  r22 = select r17, r6, r20
  p24 = select r17, p5, p23
  *@\marktext{die p24}@**@\label{line:own-oseair-die}@*
  *@\marktext{r29 = get\_cache p6}@**@\label{line:own-oseair-getcache}@*
  r30 = r29 == 42 *@\label{line:own-oseair-assert1}@*
  r31 = r29 > 43
  r32 = r30 || r31
  A = not r32  *@\label{line:own-oseair-assert2}@*
  assume A
  assert false
  halt
}
\end{lstlisting}
\end{lrbox}%

\newsavebox{\figownformula}
\begin{lrbox}{\figownformula}%
	$
		\begin{array}{l}
			\sind[1] \marktext{$p_{2}.\loc = \mathit{addr}_{0}\land{} m_0 = m_3$} \land{}   \\
			\sind[1] r_{6} = (r_{5} = 0) \land{}                                                                 \\
			\sind[1]\marktext{$ p_{3}.\loc = p_{2}.\loc \land{} p_{3}.\val = r_{5}$}  \land{}               \\
			\sind[1] \marktext{$p_{5}.\loc = p_{3}.\loc \land{} p_{6}.\loc = p_3.\loc \land{} $}     \\
			\sind[1] \marktext{$\entangle(p_{5}.\ret{val}, p_{6}.\val) \land{} p_{5}.\val = p_{3}.\val$} \land{} \\
			\sind[1] r_{17} = (r_{15} = 0) \land{}                                                               \\
			\sind[1] r_{19} = r_{18} > 1 \land{}                                                                 \\
			\sind[1] r_{20} = r_{6} \land{} r_{19} \land{}                                                       \\
			\sind[1] \marktext{$p_{23}.\loc = p_{5}.\loc \land{} p_{23}.\val = r_{18}$} \land{}             \\
			\sind[1] r_{22} = \mathit{ite}(r_{17}, r_{6}, r_{20}) \land{}                                        \\
			\sind[1] p_{24} = \mathit{ite}(r_{17}, p_{5}, p_{23}) \land{}                                        \\
			\sind[1] \marktext{$p_{24}.\ret{val} = p_{24}.val$} \land{} r_{22} \land{}                           \\
			\sind[1] \marktext{$r_{29} = p_{6}.\val$} \land{}                                                    \\
			\sind[1] r_{30} = r_{29} = 0 \land{}                                                                 \\
			\sind[1] r_{31} = r_{29} > 1 \land{}                                                                 \\
			\sind[1] r_{32} = (r_{30} \lor r_{31}) \land{}                                                       \\
			\sind[1] a = \neg r_{32} \land{}                                                                     \\
			\sind[1] a \land{}                                                                                   \\
			\sind[1] \neg \mathit{false}
		\end{array}
	$
\end{lrbox}%
\newsavebox{\figborrpd}
\begin{lrbox}{\figborrpd}%
	\begin{lstlisting}[numbers=none]
fun main() {
BB0:
  $\dots$
  P2 = mk_own P0
  *@
  \drawfatptr{p_2}{\val}{x}{\ret{\val}}{y}
  @*
  $\dots$
  p3 = set_prop P2 R5
  *@\drawfatptr{p_3}{\val}{r5}{\ret{\val}}{y}@*
  $\dots$
  P5 = mut_mkbor P3
  P6 = mut_mksuc P3
  *@
	\begin{tabular}{|c|c|c}
		\cellcolor{purple!30} $\mathit{ptr}$ & \cellcolor{purple!30} \val & \cellcolor{purple!30} \ret{\val} \\
		\cellcolor{pink!60} $\mathit{p5}$   & \cellcolor{pink!60} $\mathit{r_{5}}$   & \cellcolor{pink!60} $\boldsymbol{\mathit{z}}$ \\
    \cellcolor{orange!50} $\mathit{p6}$   & \cellcolor{orange!50} $\boldsymbol{\mathit{z}}$   & \cellcolor{orange!50} $\mathit{y}$
	\end{tabular}
  @*
  $\dots$
  br R17, ERR, BB1

BB1:
  $\dots$
  P23 = set_prop P5 R18
  *@\drawfatptr{p_{23}}{\val}{r_{18}}{\ret{\val}}{\boldsymbol{\mathit{z}}}@*
  $\dots$
  br ERR

ERR:
  $\dots$
  P24 = select R17, P5, P23
  *@\drawfatptr{p_{24}}{\val}{\mathit{ite}(r_{17}, r_5, r_{18})}{\ret{\val}}{\boldsymbol{\mathit{z}}}@*
  die P24
  *@\marktext[pink!60]{$\boldsymbol{\mathit{z}} = \mathit{ite}(r_{17}, r_5, r_{18})$}@*
  R29 = get_prop P6
  *@\marktext[pink!60]{$r_{29}=\boldsymbol{\mathit{z}}$}@*
  $\dots$
  halt
}
\end{lstlisting}
\end{lrbox}%

\begin{figure}[t]
	\begin{tabular}{l c r}
		\begin{subfigure}[t]{0.3\textwidth}
			\scalebox{1}{\usebox{\figownrpd}}
			\caption{\oseair program.}
			\label{fig:own-reduced-PD}
        \end{subfigure}
       \hspace{-1.8cm}
		\begin{subfigure}[t]{0.3\textwidth}
			\raisebox{1.7cm}{\scalebox{0.7}{\usebox{\figownformula}}}
			\caption{\smtlib program.}
			\label{fig:own-vc-code}
		\end{subfigure}
	\end{tabular}
	\caption{Program from~\cref{fig:own-example-C} in \oseair and \smtlib forms. Ownership intrinsics and their counterpart expressions in \smt are highlighted in \marktext{yellow}.}
	\label{fig:own-code-vc-translate}
	\vspace{-0.20in}
\end{figure}

\bparagraph{Verification pipeline.}
To evaluate the efficacy of ownership intrinsics for verification, we use the \seabmc bit-precise bounded model checker.
\seabmc operates on \llvm IR programs.
For this work, the \seabmc VCGen process has been enhanced to handle ownership instructions.
It is cumbersome to construct low-level \oseair programs by hand to be verified in \seabmc.
To ease the task, we provide an API for adding ownership semantics to C programs resulting in a C-like programming language with ownership semantics.
The API is in the form of C macros.
The C program is compiled to an \oseair program.
The low-level \oseair program then generates the VC in \smtlib form.
This is finally sent to an \smt solver.
We discuss the API using the example high level program in~\cref{fig:own-example-C}.
The program starts in line~\ref{line:own-example-C-mkown}, the \code{MK_OWN} macro allocates a byte of memory to an owning pointer.
The next line uses the \code{nd_char} function to assign a non-deterministic char to \code{c}.
The value of \code{c} is constrained to be \code{42} using an \code{assume} statement.
In line~\ref{line:own-example-C-setcache}, the cache at pointer \code{p} is set to the value of \code{c} using the \code{SET_CACHE} macro.
The value is also stored in memory using pointer \code{p}.
The macro \code{MUT_BORROW} in line~\ref{line:own-example-borrow} then creates a mutable borrow.
Internally, the macro expands to \code{mut_mkbor} and \code{mut_mksuc} with \code{b} getting the mutable borrow and \code{p} getting the succeeding pointer.
Next, the non-deterministic boolean value from \code{nd_bool} is used in line~\ref{line:own-example-cond} to conditionally update \code{b}'s cache to a non-deterministic value greater than \code{43}.
The \code{escapeToMemory} function takes the address of \code{b} thwarting any optimisation attempts to promote \code{b} to a register.
Finally, \code{b} dies in line~\ref{line:own-example-die} using the macro \code{DIE}.
The succeeding pointer's cache is now read using \code{GET_CACHE} into \code{r} in line~\ref{line:own-example-getcache}.
The \code{sassert} (static assert) then checks that the value of \code{r} is either \code{42} or greater than \code{43}.

For the program in~\cref{fig:own-example-C},~\cref{fig:own-reduced-PD} is its \oseair form and~\cref{fig:own-vc-code} is the generated VC.
We now describe the VCGen in \machtwo using~\cref{fig:own-code-vc-translate}.
The high-level ownership instructions are highlighted in yellow in both figures.
The \code{MK_OWN} macro in C becomes the \code{mk_own} instruction in \oseair and is translated to \smtlib form using \symfun. Note that in \machtwo, \code{mk_own} does not write to memory or update the pointer cache.
The symbolic semantics therefore only allocates memory and provides a previously unallocated address $addr_{0}$.
The \code{set_cache} instruction in line~\ref{line:own-oseair-setcache} defines a pointer $p_{3}$ with the same \loc as $p_{2}$ and the cache updated to $r_{5}$.
The mutable borrow occurs in lines~\ref{line:own-oseair-mkbor}--\ref{line:own-oseair-mksuc}.
The semantics copies the lender $p_{3}.\loc$ to $p_{5}.\loc$ and $p_{6}.\loc$.
The \val and \ret\val fields are set up as usual.
The cache of the borrowed pointer $p_{5}$ is conditionally updated in line~\ref{line:own-oseair-setcache2}.
The borrowed (variant) pointer $p_{24}$ dies in line~\ref{line:own-oseair-die} with the usual semantics.
The cache of the succeeding pointer $p_{6}$ is read into $r_{29}$ in line~\ref{line:own-oseair-getcache}.
The lines~\ref{line:own-oseair-assert1} to~\ref{line:own-oseair-assert2} set up the program such that if an execution satisfies \code{assume A} then it reaches the error state (\code{assert false}).
An important consequence of using ownership semantics is that the \smtlib program does not need to model the \code{store} instruction in line~\ref{line:own-oseair-store}.

\section{Evaluation}
\label{sec:eval}
\newsavebox{\figproof}
\begin{lrbox}{\figproof}%
  \begin{lstlisting}[language=C, escapechar=@, numbers=left]
      enum status {O, C};
      int unit_proof(const char **fnames,
          int n) {
        FILE *f[MAX];// assume n < MAX
        for(int i=0; i < n; i++) {
          set_shad(f[i], O);
          f[i] = open(fnames[i], "w");}
        size_t choose = nd_size_t();
        assume(choose < n);
        FILE *file = f[choose];
        write(file);
        // check file closed
        sassert(get_shad(file) == C);}
 \end{lstlisting}
\end{lrbox}

\newsavebox{\figsut}
\begin{lrbox}{\figsut}%
	\begin{lstlisting}[language=C, escapechar=@, numbers=left]
    void write(FILE *fp) {
      // check file opened
      sassert(get_shad(fp) == O);
      fputc('a', fp);
      // mark closed
      set_shad(fp, C);
      fclose(fp);}
 \end{lstlisting}
\end{lrbox}
\begin{figure}[t]
\centering
	\subcaptionbox{A unit proof.\label{fig:proof_example}}[0.55\linewidth]{%
		{\scalebox{1.0}{\usebox{\figproof}}}}
	\subcaptionbox{An SUT.\label{fig:sut_example}}[0.4\linewidth]{%
		{\scalebox{1.0}{\usebox{\figsut}}}}
	\caption{An example of typestate storage in shadow memory.}
    \label{fig:typestate_example}
\vspace{-0.25in}
\end{figure}
\begin{figure}[t]
    \centering
    \includegraphics[width=0.5\textwidth]{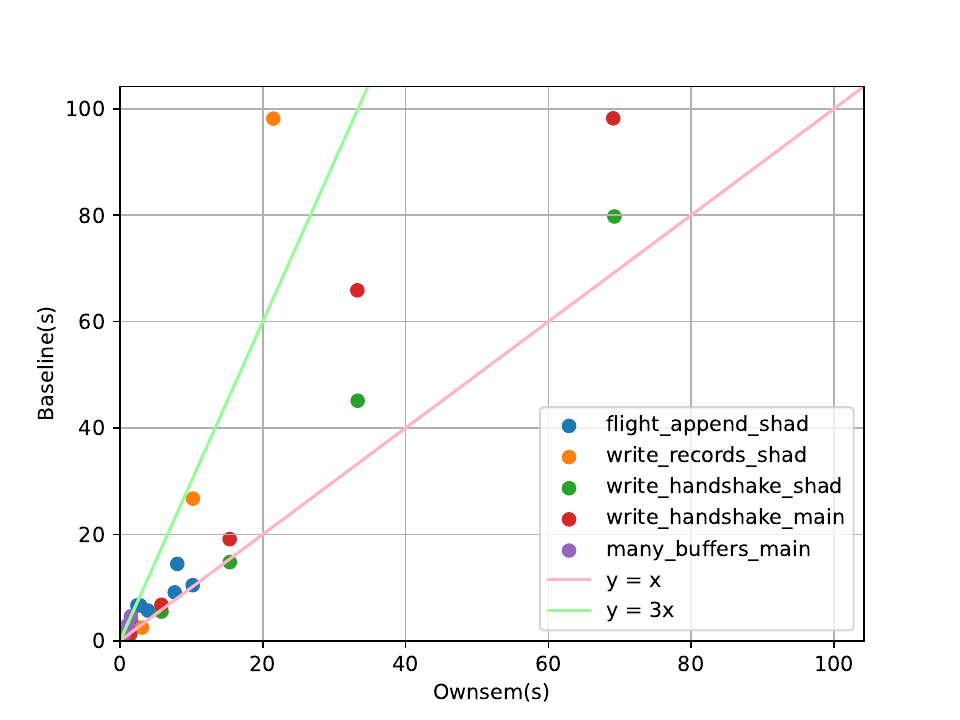}
    \caption{Solve time (in sec.) using ownership semantics vs baseline.}
    \label{fig:time_scatter}
\vspace{-0.25in}
\end{figure}
We would like to cache verification properties for practical programs.
We first describe properties of interest and our baseline property carrying mechanisms through the example in~\cref{fig:typestate_example}.
Here, we want to check that a \code{write} function in ~\cref{fig:sut_example} writes only to an \emph{open} file and the file is always \emph{closed} after a write.
The state of the file object is encoded as a \emph{typestate} property~\cite{6312929} that records (and checks) operations that have occurred on an object.
The \code{write} function is called using the \emph{unit proof} harness in~\cref{fig:proof_example}.
It defines \code{n} file pointers based on user input.
The typestate is marked open for each file pointer in \emph{shadow memory} using the \code{set_shad} function.
Shadow memory is an address map ($\loc$ to \bv{64}).
It is used to stash verification relevant object metadata.
In practice, shadow memory is provided by verification and testing tools like \seabmc and Memcheck~\cite{DBLP:conf/vee/NethercoteS07}.
If shadow memory is not available then metadata can be stashed in a separate main (program) memory allocation, which is available when the program is built in debug or verification mode.
In the example, after the typestate is set to open for each file pointer, we choose one file pointer \code{file} out of \code{n} for calling \code{write} --- the system under test (SUT).
In the SUT, we first check that the typestate of the file object is open using \code{get_shad} to get the typestate value.
Then the char \code{'a'} is written and the file is closed, with the typestate marked as closed.
Finally, the harness checks that the file typestate is indeed closed.

Note in the given unit proof, the write and read of shadow memory can resolve to one of \code{n} allocations since any file object can be chosen.
Therefore, in the VC, memory access would involve solving an ITE (if-then-else) expression for 1 of n choices.
Here, solving the ITE is redundant since we only want to check that a given operation occurred on the chosen file pointer.
A better solution would be to store typestate in the file (fat) pointer cache itself.
With this optimisation, an ITE would not be traversed since read, writes of the typestate would be at the pointer (using \code{get_cache} and \code{set_cache}) leading to simpler VC.

We base our experiments on this idea utilizing the \seabmc model checking engine for \seahorn.
\seabmc originally takes a \seair program as input and generates VC that are solved by an SMT solver.
We enhance \seabmc to now take \oseair programs as input.
Using the C macro API, C unit proofs are compiled to \smt.
We then measure how typestate cached at pointers compares to typestate stored in memory.

The C unit proofs we work with come from \mbedtls~\cite{mbedTLS}, a C library of cryptographic primitives, SSL/TLS and DTLS protocols.
In particular, we look at three functions in \code{ssl_msg.c} that handles SSL message construction and de-construction.
The flow we consider are
\begin{inparaenum}[(1)]
  \item \code{flight_append} that appends messages to the current flight of messages,
  \item \code{write_records} that encrypts messages into records and sends them on the wire, and,
  \item \code{write_handshake} that writes handshake messages.
\end{inparaenum}
Each SUT operates on a byte buffer data structure.
We are interested in recording and checking typestate properties for such a buffer.
However, similar to example~\cref{fig:typestate_example}, the unit proof is set up such that a single byte buffer pointer may point to 1-of-n buffer objects.
Therefore, we study if using pointer caching improves solver performance.

The experiments are run on an Intel(R) Xeon(R) E5-2680 CPU operating at 2.70GHz with 64 GiB of main memory.
The generated VC are solved using Z3~\cite{DBLP:conf/tacas/MouraB08} \code{smtfd} tactic.
The scatter plot in~\cref{fig:time_scatter} shows the solving time for unit proof with ownership semantics (ownsem) in the x-axis.
The y-axis records the solving time for the same unit proof that either uses shadow memory or main memory as the baseline.
The legend makes it clear which memory we compare against using either \emph{shad} or \emph{main} in the name suffix.
We run each flow for increasing number of byte buffers behind a pointer (e.g., 2, 4, 6, \ldots) and stop when the running time in either ownsem or baseline mode reaches 100 seconds.
The \code{many_buffers} benchmark is a handcrafted benchmark that shows a consistent 3x improvement for ownsem.
The flows from \mbedtls show more spread.
For small number of buffers, ownsem and baseline are usually head-to-head.
As the number of buffers increase, ownsem outperforms baseline.
For \code{write_handshake_shad}, the performance boost is 1.3x when using 8 buffers.
For \code{write_records_shad}, the performance boost at 8 buffers is 5x.
This is shown on the scatter plot.
Looking at the \smt solver metrics, we see internal metrics like \code{sat conflicts} and \code{sat backjumps} correlate with the timings (See%
\ifarxivmode
~\cref{app:appendixb}
\else
~\cite{priya2024ownershiplowlevelintermediaterepresentation}
\fi
for details).
When a unit proof is faster, then it sees fewer conflicts and backjumps compared to the baseline.
This is indirect evidence that the performance boost is due to simplicity of the VC.

Simplification of VC itself depends on how many memory accesses can be soundly replaced by pointer cache accesses.
VC simplicity is affected by
\begin{inparaenum}[(1)]
  \item the extent a conditional typestate check depends on program memory state, and
  \item the number of typestate memory accesses as a fraction of the total number of memory accesses.
\end{inparaenum}
 As an example, for a conditional check such as \code{if(*unrelated_ptr == 1) \{get_cache(ptr);\}},  a read of \code{ptr} cache using \code{get_cache} does not access \code{ptr} memory. However, for the check to be reachable, the guarding \code{if} condition does require a memory access.
Therefore, it is not always possible to remove dependency on program memory for conditional typestate checks.
Also if the unit proof (and the SUT) do not set/get typestate checks frequently then replacing such checks by pointer cache accesses has limited benefits.
The data and units proofs to reproduce our experiments are available at \url{https://github.com/priyasiddharth/mbedtls-ownsem}.

Overall, a speedup in solving time occurs as expected.
The speedup is due to simpler VC.
However, the speedup is sensitive to the property expressed as a typestate check and number of operations on object (pointer) that affect typestate.

\section{Related work}
\label{sec:related}
 RustBelt~\cite{DBLP:journals/pacmpl/0002JKD18}, Oxide~\cite{DBLP:journals/corr/abs-1903-00982} formalize subsets of high level Rust.
RustBelt uses a continuation passing style functional language to describe the semantics.
Oxide uses a high level language similar to \rust.
These approaches do not map directly to a low--level register machine like \llvm.
Stacked Borrows~\cite{DBLP:journals/pacmpl/JungDKD20} formulates \rust ownership semantics as a stack discipline working on de--sugared (MIR) \rust syntax that represents memory by an address map.
Its aim is to provide a reference semantics for the borrow checker separate from the production version in the \rust compiler.
Stacked Borrows is implemented in the MIR interpreter (MIRI) and is part of the \rust standard distribution.
We rely heavily on stacked borrows to design low-level semantics for this paper.

The Move Prover~\cite{DBLP:conf/tacas/DillGPQXZ22} uses reference elimination to replace a reference by its data.
It assumes an alias free memory model and solves the problem of return of a mutable borrow by recording the origin (lender) of a mutable borrow and returning data to it explicitly rather than utilizing prophecies.
RustHorn~\cite{DBLP:journals/toplas/MatsushitaTK21} uses a prophecy value to model return of a mutable borrow and assumes a safe \rust-like language and, therefore, forgoes modelling an address map entirely.
RustHornBelt~\cite{DBLP:conf/pldi/MatsushitaDJD22} extends this work to cover unsafe \rust where the safety in the unsafe part is manually proven in Iris~\cite{DBLP:conf/popl/JungSSSTBD15}, a concurrent separation logic prover built on top of Coq~\cite{Coq}.

Verus~\cite{DBLP:journals/pacmpl/LattuadaHCBSZHPH23}, Prusti~\cite{DBLP:conf/nfm/AstrauskasBFGMM22}, and Creusot~\cite{DBLP:conf/icfem/DenisJM22} are deductive verifers for \rust. Creusot uses RustHorn style prophecy variables.
These deductive tools can model complicated features of the language, like polymorphism, directly.
This paper focuses on low-level memory manipulating programs.

The memory models used in CBMC~\cite{DBLP:journals/corr/abs-2302-02384}, LLBMC~\cite{DBLP:conf/vstte/MerzFS12}, and stock SeaBMC~\cite{DBLP:conf/fmcad/PriyaSBZVG22} assume an unsafe language allowing unrestricted aliasing of pointers and support pointer arithmetic.
Kani~\cite{KaniVerifier} is a \rust verifier that compiles to goto-cc, the same low-level backend as CBMC.
Ownership information, though, is lost is this conversion.
Overall, we expect these low-level tools would perform similar to our baseline experiments.

\section{Conclusion}
\label{sec:conclusion}
We describe formal ownership semantics for multiple low-level machines of increasing complexity.
Particularly, we explain the mechanism for caching values at fat pointers and keeping the values in-sync with memory.
We use the given semantics to describe VCGen for \bmc
such that the number of occurrences of memory accesses in the VC is reduced.
For this we model return of mutable borrows using prophecy values added to fat pointers.
We evaluate the efficiency of generated VC by experiments using the \seabmc tool.
Overall, we see improvements in solving time and attribute it to the simplicity of VC.

\bibliographystyle{IEEEtran}
\bibliography{bib}
\ifarxivmode
	\appendix
	\subsection{Appendix A}
\label{app:appendixa}

\begin{table*}[t]
	\centering
	\renewcommand{\arraystretch}{1.5} 
	\begin{tabular}{l|l|l}
		Operation                                                                        & Pre-condition & Post-condition                                                                                                                                                                                                                                                                                                           \\
		\hline
		\code{die q}                                                                     &
		\begin{tabular}{@{}l@{}}
			$R[\mathsf{q}]=(q.\loc,tag_q,n_q)$
		\end{tabular}                               &
		\begin{tabular}{@{}l@{}}
			$(\exists p.R[p]=(q.\loc,tag_p,\_) \implies R[p]=(q.\loc,tag_{p},n_{q})),$ \\
			$(\exists a.M[a]=(q.\loc,tag_p,\_) \implies M[a]=(q.\loc,tag_{p},n_{q}))$
		\end{tabular}                                                                                                                                                                                                                                                                                                  \\
		\hline
		\code{m1 = store r, p, m0}                                                       &
		\begin{tabular}{@{}l@{}}
			$R[\mathsf{p}]=(p.\loc,tag_p,\_)$
		\end{tabular}                                &
		\begin{tabular}{@{}l@{}}
			$R[\mathsf{p}]=(p.\loc,tag_{p},v_{s}),M[p.\loc] = v_{s},$ \\
			$(\mathsf{r} = v) \implies (v_{s} = v),$                  \\
			$(\mathsf{r} = (tag,t,v)) \implies (v_{s} = (tag,t))$
		\end{tabular}                                                                                                                                                                                                                                                                                                                         \\
		\hline
		\code{r = load p, m}                                                             &
		\begin{tabular}{@{}l@{}}
			$R[\mathsf{p}]=(p.\loc, tag_p, v_{p})$
		\end{tabular} &
		\begin{tabular}{@{}l@{}}
			$R[\mathsf{r}]=v_{r},$                                     \\
			$R[\mathsf{p}]=(p.\loc, tag_p, v_{r}),$                    \\
			$(B_0 = \varnothing, t_{p} \in\{\mathsf{o},\mathsf{mb}\},$ \\$\quad((\exists \loc_{p_{1}} \exists tag_{p_{1}}.(\loc_{p_{1}},tag_{p_{1}})=v_p,v_{p_{1}} = M[\loc_{p_{1}}]) \implies (v_{r} = (\loc_{p_{1}},tag_{p_{1}}, v_{p_{1}}))),$\\
			$\quad(\exists v .v = v_p) \implies (v_{r} = v)),$         \\
			$(t_{p}=\mathsf{c} \implies (v_r = M[p.\loc])$
		\end{tabular} \\
	\end{tabular}
	\caption{Effect of selected operations on $S_{B}$,$R$, and $M$ in machine \machthree. The remaining instructions stay the same as~\cref{tab:effect_m1}.}
	\label{tab:effect_m3}
\end{table*}
\bparagraph{Semantics of \machthree.}
The machines \machzero and \machone do not support storage of fat pointers to memory.
We now define an extension to \machone called \machthree, that allows this in~\cref{tab:effect_m3}.
We only show definitions for instructions that change from \machone.
A \code{die} instruction returns a mutable borrow as before.
However, since the succeeding pointer may exist in the register map or memory map, the search for $tag_p$ now occurs in both maps.
Once the succeeding pointer is found, the pointer cache is updated as before to $n_q$.

A \code{store} and \code{load} operation now becomes more involved.
This is because when storing a fat pointer in memory, the memory must be wide enough to contain $(\loc,tag,\val)$ fields.
However, the \val field may itself be a pointer, leading to unbounded recursion.
We work around this by bounding the memory to be wide enough to contain $(\loc,tag,\val)$ where \val is word size.
A \code{store} of a value $r$ using pointer-to-pointer $p$ has two cases.
If $r$ is a (scalar) value then it is stored (and cached) as in \machone.
If $r$ is a fat pointer then we remove the $r.\val$ field to make a value of type $(r.\loc,r.tag)$.
This value is then written both to memory and to $p.\val$.
Therefore, the pointer cache now contains either a scalar or a tuple $(r.\loc,r.tag)$.

A \code{load} has three distinct behaviours.
First, if the lender pointer $p$ is
\begin{inparaenum}[(1)]
	\item top of borrow stack, and
	\item mutably borrowed or owning, and
	\item a bit-vector (scalar) value, then
\end{inparaenum}
the read from memory is replaced by a read of the $p.\val$ field.
Second, if the \code{load} uses a lender pointer $p$ that is
\begin{inparaenum}[(1)]
	\item top of borrow stack, and
	\item mutably borrowed or owning, and
	\item a partial pointer value $(p.\loc,p.tag)$, then
\end{inparaenum}
the read from memory is replaced by a read of the $p.\val$ (cache) field.
This is used to write $(r.\loc, r.tag)$
Additionally, memory is read at $M[\loc]$ to get the $r.\val$ field.
The construction of the value $(r.\loc,r.tag,r.\val)$ is now complete.
In the third case, the \code{load} uses a copied lender pointer $p$ and value is read directly from memory.
Note that in this case, if $r$ is a pointer then its cache will contain a garbage value.
With these changes, \machthree is a practical machine that can \code{load} and \code{store} pointers.

\subsection{Appendix B}
\label{app:appendixb}
\begin{figure}[t]
	\centering
	\includegraphics[width=0.5\textwidth]{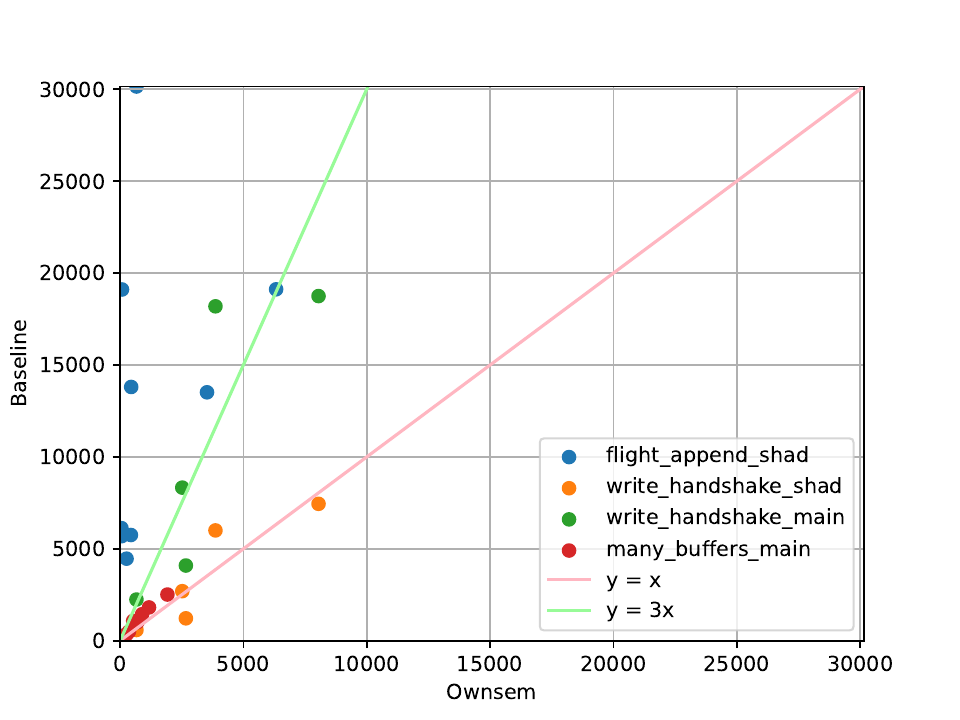}
	\caption{Number of SAT conflicts using ownership semantics vs baseline.}
	\label{fig:conflicts_scatter}
	\vspace{-0.25in}
\end{figure}

\begin{figure}[t]
	\centering
	\includegraphics[width=0.5\textwidth]{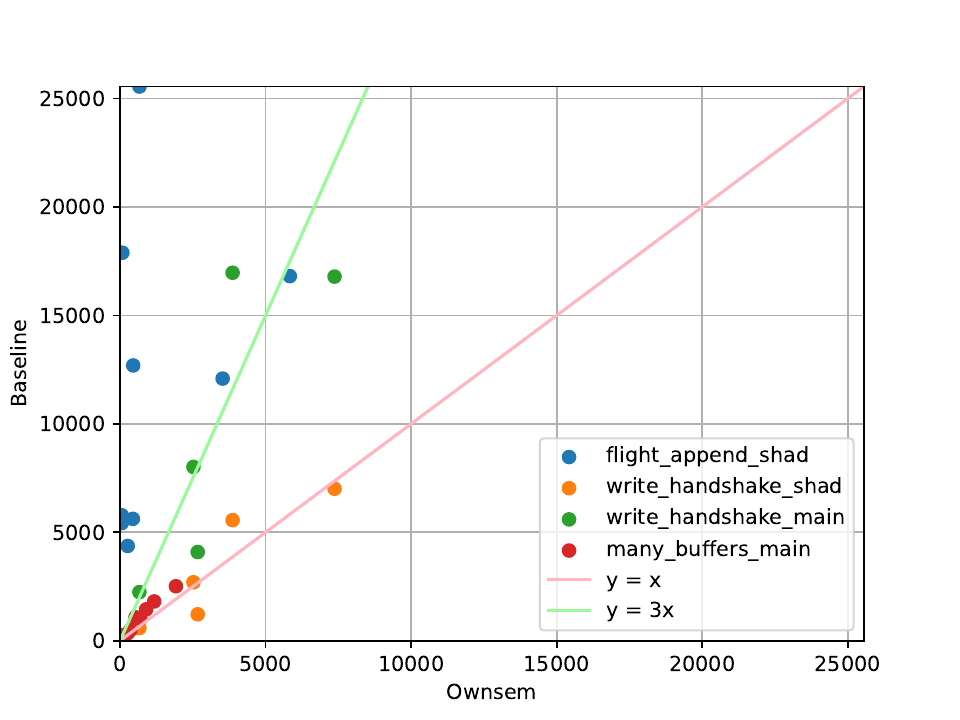}
	\caption{Number of SAT backjumps using ownership semantics vs baseline.}
	\label{fig:backjumps_scatter}
	\vspace{-0.25in}
\end{figure}

Solving time gives an overall measure of how caching at pointer (ownsem) compares to memory (baseline).
For evidence that ownsem problems are easier for the SMT solver, we look at the number of conflicts in (\cref{fig:conflicts_scatter}) and number of backjumps (\cref{fig:backjumps_scatter}) in both settings.
The dots correspond to different number of buffer as in~\cref{sec:eval}.
We can see the number of conflicts in ownsem is generally much lower than baseline configurations.
The \code{write_handshake_shad} flow is an exception and, correspondingly, the speedup obtained is lowest in this case (1.3x).
The \code{write_records_shad} ownsem flow reports only one conflict consistently for all buffer sizes vs a large number of conflicts (around 500k) for baseline for 8 buffers.
Including this metric would skew the graph towards larger numbers hiding details, hence we exclude it.
The number of backjumps follows a similar trajectory where the number of backjumps seen for ownsem is lower than baseline with the exception of \code{write_handshake_shad}.
The \code{write_records_shad} flow again shows lower number of backjumps for ownsem vs baseline. We do not include it so as to not skew the graph towards larger numbers.
Overall, both the number of conflicts and the number of backjumps correlate strongly with faster solving time.

\fi
\end{document}